\documentclass[final,3p,times]{elsarticle}

\usepackage{graphicx}

\usepackage{amsmath}
\usepackage{graphicx,bm}
\usepackage{ulem}
\usepackage{amssymb}
\usepackage{color}
\usepackage{version}

\usepackage[english,francais]{babel}

\def\og{\leavevmode\raise.3ex\hbox{$\scriptscriptstyle\langle\!\langle$~}}
\def\fg{\leavevmode\raise.3ex\hbox{~$\!\scriptscriptstyle\,\rangle\!\rangle$}}

\usepackage[T1]{fontenc}
\usepackage{amsmath}
\usepackage{color}
\newcommand{\rr}{\mathbf{r}}
\newcommand{\kk}{\mathbf{k}}
\newcommand{\qq}{\mathbf{q}}

\newcommand{\vv}{\mathbf{v}}
\newcommand{\dd}{\mathrm{d}}
\newcommand{\ii}{\mathrm{i}}
\newcommand{\eee}{\mathrm{e}}

\newcommand{\bra}[1]{\langle #1|}
\newcommand{\ket}[1]{|#1\rangle}
\newcommand{\meanv}[1]{\langle #1 \rangle}
\newcommand{\meanvlr}[1]{\left\langle #1 \right\rangle}

\newcommand{\bb}[1]{\left( #1 \right)}
\newcommand{\bbcro}[1]{\left[ #1 \right]}
\newcommand{\be}{\begin{equation}}
\newcommand{\ee}{\end{equation}}
\newcommand{\bea}{\begin{eqnarray}}
\newcommand{\eea}{\end{eqnarray}}
\newcommand{\beq}{\begin{equation}}
\newcommand{\eeq}{\end{equation}}
\newcommand{\g}[1]{\og#1\fg}

\newcounter{notecond}
\newcounter{noteansatz}
\newcounter{notepaire}
\newcounter{notetauc}
\newcounter{notetaucexpli}

\newcommand{\yvan}{}
\newcommand{\gilles}{}

\usepackage{float}

\begin{document}
\selectlanguage{english}

\begin{frontmatter}


\selectlanguage{english}
\title{Thermal blurring of a coherent Fermi gas\\
}


\author{Hadrien Kurkjian}
\author{Yvan Castin}
\author{Alice Sinatra}

\address{Laboratoire Kastler Brossel, ENS-PSL, CNRS, UPMC-Sorbonne Universit\'es et Coll\`ege de France, Paris, France}



\begin{abstract}
It is generally assumed that a condensate of paired fermions at equilibrium is characterized by a macroscopic wavefunction with a well-defined, immutable phase. In reality, all systems have a finite size and are prepared at non-zero temperature; the condensate has then a finite coherence time, even when the system is isolated in its evolution and the particle number
$N$ is fixed. The loss of phase memory is due to interactions of the condensate with the excited modes that constitute a dephasing environment. This fundamental effect, crucial for applications using the condensate of pairs' macroscopic coherence, was scarcely studied. We link the coherence time to the condensate phase dynamics, and we show with a microscopic theory that the time derivative of the condensate phase operator $\hat{\theta}_0$ is proportional to a chemical potential operator
that we construct including both the pair-breaking and pair-motion excitation branches.  In a single realization of energy $E$,
$\hat{\theta}_0$ evolves at long times as $-2\mu_{\rm mc}(E)t/\hbar$ where $\mu_{\rm mc}(E)$ is the microcanonical chemical potential; energy fluctuations from one realization to the other then lead to a ballistic spreading of the phase and to a Gaussian decay of the temporal coherence function with a characteristic time $\propto N^{1/2}$.
In the absence of energy fluctuations, the coherence time scales as $N$ due to the diffusive motion of $\hat{\theta}_0$.
We propose a method to measure the coherence time with ultracold atoms, which we predict to be tens of milliseconds for the canonical ensemble unitary Fermi gas.

\noindent 
\vskip 0.5\baselineskip

\selectlanguage{english}
\end{abstract} 
\end{frontmatter}

\selectlanguage{english}


\section{Setting the stage}

Coherent gases of ultracold atoms confined in immaterial non-dissipative traps are unique examples of isolated macroscopic quantum systems. The value of their intrinsic coherence time is therefore a fundamental question. But it is also a practical issue for all the applications which exploit macroscopic coherence, such as interferometry 
or quantum engineering where one generates non-trivial entangled states by coherent evolution \cite{Bloch2002_Revival,Oberthaler2008,Treutlein2010}.
Coherence time measurements are presently being performed in cold Bose gases \cite{Ketterle2007,Sidorov2011,Berrada2013}. Experiments on Fermi gases, which up to now focused on traditional aspects of the $N$-body problem, such as thermodynamic properties \cite{Salomon2010,Zwierlein2012}, are moving towards correlation and coherence measurements \cite{Kohstall2011}. This turn will open a new research field, including the strong coupling regime : that of fermionic quantum optics \cite{CarusottoCastin2005}.
However, a theory predicting the coherence time of  a pair-condensed Fermi gas was missing, except in
the limiting case of zero temperature \cite{PRA2013}. In this paper we present the first microscopic theory bridging this
theoretical gap in a general way. Our approach holds for other physical systems, such as mesoscopic Josephson Junctions, provided that the environment-induced decoherence is sufficiently reduced.

For a Bose-condensed gas of bosons, the finite coherence time is due to the spreading of the probability distribution
{\gilles of the condensate phase change during $t$}. At zero temperature and in presence of interactions, a ballistic phase spreading is caused by atom number fluctuations in the sample. This effect has been measured by interfering two initially mutually-coherent condensates, whose particle number fluctuates due to partition noise \cite{Ketterle2007,Berrada2013}. Contrarily to lasers, which are open quantum systems, and somehow unexpectedly, a ballistic spreading persists in Bose-Einstein condensates for a fixed atom number at non-zero temperature \cite{Sinatra2007,Kuklov2000}.  Fluctuations of the energy, another conserved quantity, then play the same role as number fluctuations.

For an unpolarized pair-condensed Fermi gas, the study of coherence time presupposes a clear definition of the condensate phase and of the corresponding operator $\hat{\theta}_0$ \cite{PRA2013}. Furthermore, at non-zero temperature
the speed of variation of the phase should include the contribution of two excitation branches: the fermionic pair-breaking one and the bosonic one exciting the pair motion. For the fermionic branch Anderson's Random Phase Approximation (RPA) \cite{Anderson1958} is enough. For the bosonic branch however, we need the equivalent for  fermions of the Bogoliubov method to construct quasiparticle creation 
$\hat{b}_\alpha^\dagger$ and annihilation $\hat{b}_\alpha$ operators and to express $\dd \hat{\theta}_0/\dd t$  in term of these operators. More than that, we need to include interactions among quasiparticles in the evolution of the  
$\hat{b}_\alpha$. This is a non-trivial consequence of the dependence of condensate wavefunction on the 
total number of particles $N$ even for a spatially homogeneous system, and clearly goes beyond the RPA program. 

\section{Correlation function decay} 
Below the critical temperature, the time-correlation function of the pairing field
$\hat{\psi}_\downarrow(\rr) \hat{\psi}_\uparrow (\rr')$ where $\hat{\psi}_\sigma(\rr)$ is the fermionic field operator of the spin
$\sigma$ component, is dominated at long times by the condensate contribution:
\begin{equation}
g_1(t)=\meanv{\hat{a}_0^\dagger(t) \hat{a}_0(0)}
\label{eq:g1}
\end{equation}
where
$
\hat{a}_0 = \int \dd^3 r \dd^3 r' \varphi_0^{\gilles *}(\rr,\rr') \hat{\psi}_\downarrow(\rr) \hat{\psi}_\uparrow(\rr')
$
is the component of the pairing field on the condensate wavefunction \cite{PRA2013}. At equilibrium the system is in a mixture of $N$-body eigenstates $\ket{\psi_\lambda}$, with weights $\Pi_\lambda$. We therefore study the correlation function
 $g_1^\lambda(t)$ in the eigenstate $\ket{\psi_\lambda}$ of energy $E_\lambda$ and particle number $N_\lambda$.
To exploit the weak relative fluctuations in the number of condensed pairs for a large system, we split $\hat{a}_0$
into modulus and phase hermitian operators \cite{PRA2013} 
\begin{equation}
\hat{a}_0=\eee^{\ii \hat{\theta}_0} \hat{N}_0^{1/2},
\label{eq:deftheta0}
\end{equation} 
and we approximate {\yvan the number of condensed pairs operator} $\hat{N}_0$ by its mean value $\bar{N}_0$ in the equilibrium state to obtain
\begin{equation}
g_1^\lambda(t)\simeq \bar{N}_0 \eee^{\ii E_\lambda t /\hbar} \bra{\psi_\lambda} \eee^{-\ii (\hat{H}+\hat{W})t/\hbar} \ket{\psi_\lambda}
\label{eq:g1lam}
\end{equation}
The  operator $\hat{W}$, difference between $\hat{H}$ transformed by $\eee^{\ii\hat{\theta}_0}$ and $\hat{H}$,
\begin{equation}
\hat{W}=\eee^{-\ii\hat{\theta}_0} \hat{H} \eee^{\ii\hat{\theta}_0} - \hat{H}=- \ii [ \hat{\theta}_0 , \hat{H} ]-\frac{1}{2} [ \hat{\theta}_0 ,[ \hat{\theta}_0 , \hat{H} ]]+\ldots
\label{eq:BKHpourW}
\end{equation}
is approximatively $N$ times smaller than $\hat{H}$. Indeed $\eee^{\ii\hat{\theta}_0}$, like $\hat{a}_0$, changes the total particle number by a quantity $O(N^0)$. While $\hat{H}$ is an extensive observable, $\hat{W}$ is intensive {and the double commutator in (\ref{eq:BKHpourW}) is of order $1/N$}.
In equation  (\ref{eq:g1lam}) formally appears the evolution operator of the Hamiltonian $\hat{H}$ perturbed by $\hat{W}$, and restricted to the eigenstate $\ket{\psi_\lambda}$ of $\hat{H}$. Up to a phase factor, the function $g_1^\lambda/\bar{N}_0$ is then  proportional to the probability amplitude that the system prepared in $\ket{\psi_\lambda}$ is still in that state after a time $t$ {\yvan in the presence of the perturbation $\hat{W}$}. A standard way to obtain a non-perturbative approximation of this amplitude is to use the Green function or the resolvent
operator $\hat{G}(z)=\bb{z\hat{1}-(\hat{H}+\hat{W})}^{-1}$ of the perturbed Hamiltonian. Within the projectors method {(see \S III.B.2 of \cite{Cohen})}, we introduce an effective non hermitian Hamiltonian  $\hat{H}_{\rm eff}(z)$ governing the evolution restricted to $\ket{\psi_\lambda}$, {\it id est} $\bra{\psi_\lambda} \hat{G}(z) \ket{\psi_\lambda}=\bb{z-\bra{\psi_\lambda} \hat{H}_{\rm eff}(z) \ket{\psi_\lambda}}^{-1}$. 
{\yvan This leads to
\be
g_1^\lambda(t) \simeq \bar{N}_0 \int_C \frac{\dd z}{2i\pi} \frac{\eee^{-\ii (z-E_\lambda) t/\hbar}}{z-\langle \hat{H}_{\rm eff}(z)
\rangle_\lambda} \ \ \ \mbox{with} \ \ \
\langle \hat{H}_{\rm eff}(z)\rangle_\lambda=E_\lambda+\langle\hat{W}\rangle_\lambda
+\langle \hat{W} \hat{Q}_\lambda \frac{\hat{Q}_\lambda}{z \hat{Q}_\lambda -\hat{Q}_\lambda (\hat{H}+\hat{W}) \hat{Q}_\lambda}
\hat{Q}_\lambda \hat{W}\rangle_\lambda
\label{eq:expliG}
\ee
The integration domain $C$ in the complex plane is any straight line $z=x+i\eta$, $\eta>0$ fixed,
where the real number $x$ runs from $+\infty$ to $-\infty$. The notation 
$\meanv{\hat{A}}_\lambda\equiv \bra{\psi_\lambda} \hat{A} \ket{\psi_\lambda}$ was used, and the 
operator $\hat{Q}_\lambda=\hat{1}-\ket{\psi_\lambda} \bra{\psi_\lambda}$, that projects orthogonally to $\ket{\psi_\lambda}$,
was introduced.}
Keeping in {\yvan $\langle\hat{H}_{\rm eff}(z)\rangle_\lambda$}
terms up to order two in $\hat{W}$, {\yvan hence omitting $\hat{Q}_\lambda \hat{W} \hat{Q}_\lambda$
in the denominator}, and neglecting the $z$ dependence 
{\yvan $\langle\hat{H}_{\rm eff}(z)\rangle_\lambda \approx \langle\hat{H}_{\rm eff}(E_\lambda+\ii 0^+)\rangle_\lambda$}
(pole approximation), we obtain\footnote{The pole approximation implicitly assumes that $|\psi_\lambda\rangle$ is coupled to a broad energy continuum {\gilles \cite{Cohen}}. As a consequence {\yvan the exponential law in}
 (\ref{eq:carrebis}) holds only at times longer than the inverse frequency width of the continuum, {\it i.e.\,} longer than the quasi particle correlation time $\tau_c$ introduced below.} \setcounter{notetauc}{\thefootnote}
\footnote{{\yvan In order to have a branch cut in the resolvent and a pole in its analytic continuation, which is required
to justify rigorously the existence of a nonzero imaginary part $\gamma_\lambda$ and to give a precise meaning to Eq.~(\ref{eq:carrebis}),
one must take the thermodynamic limit ($N\to +\infty$ for a fixed density). We face here an unusual subtlety:
as shown by a generalisation of the reasoning around Eq.~(\ref{eq:corr_theta}), the shift function \cite{Cohen}}
\yvan{$\Delta_\lambda(z)\equiv \langle \hat{W} \hat{Q}_\lambda\frac{\hat{Q}_\lambda}
{\hat{Q}_\lambda z +\hat{Q}_\lambda (E_\lambda-\hat{H})\hat{Q}_\lambda}\hat{Q}_\lambda \hat{W}\rangle_\lambda$,
with $\mathrm{Im}\, z>0$, tends to zero as $1/N$, and so does $\delta_\lambda-\ii\gamma_\lambda$ in (\ref{eq:carrebis}).
The way out is to introduce a time scale of order $N$, setting $t=N\tau$ with $\tau$ fixed at the thermodynamic limit.
One performs the change of variable $z=E_\lambda+\langle \hat{W}\rangle_\lambda + z'/N$ in the integral of (\ref{eq:expliG})
and one chooses $\eta=\eta'/N$ ($\eta'>0$ fixed) in the integration domain $C=\{z=x+i\eta,x=+\infty \to -\infty\}$,
which leads to the integration domain $C'=\{z'=x'+i\eta',x'=+\infty \to -\infty\}$ over $z'$,
and one neglects the term $\langle\hat{W}\rangle_\lambda$ in the denominator of the shift function, as well as the term $\mathrm{Re}\, z'/N$
(to be consistent with the omission of $\hat{Q}_\lambda \hat{W}\hat{Q}_\lambda$).
The factor $1/N$ in the element of integration $\dd z=\dd z'/N$ allows one to pull out a factor $1/N$ in the denominator of the integrand,
and one obtains in the thermodynamic limit the following result, 
\[
\lim_{N\to +\infty} \frac{g_1^\lambda(t) \eee^{\ii \langle \hat{W}\rangle_\lambda t/\hbar}}{\bar{N}_0} \simeq
\int_{C'} \frac{\dd z'}{2\ii \pi} \frac{\eee^{-\ii z' \tau/\hbar}}{z'-\hbar\Omega_\lambda} = \eee^{-\ii \Omega_\lambda \tau} \ \ \mbox{with}\ \
\hbar\Omega_\lambda = \lim_{N\to +\infty} N \Delta_\lambda(\ii\eta'/N)
\]
which is $\eta'$-independent due to the analyticity of the integrand in the half-plane $\mathrm{Im}\, z'>0$.
For a large but finite size system, this leads to the definition $N(\delta_\lambda-\ii\gamma_\lambda) \equiv \Omega_\lambda$.
If one rather takes the thermodynamic limit at fixed $t$ {\sl before} setting $t$ to infinity, one obtains,
after expansion of the integrand of (\ref{eq:expliG}) to first order in the shift function, {\sl another} definition:
$\lim_{N\to +\infty} N [\bar{N}_0^{-1} g_1^\lambda(t) \eee^{\ii \langle \hat{W}\rangle_\lambda t/\hbar} - 1]
\sim  (-\ii t) N(\delta_\lambda-\ii\gamma_\lambda)$
with $N(\delta_\lambda-\ii\gamma_\lambda)\equiv\lim_{\eta\to 0^+} \lim_{N\to +\infty} N \Delta_\lambda(\ii \eta)$.
To show the equivalence of these two definitions, let us inject in the shift function a closure relation on the eigenstates
of $\hat{H}$, $\Delta_\lambda(z)=\sum_{\mu\neq \lambda} |\langle\psi_\mu|\hat{W}|\psi_\lambda\rangle|^2/(z+E_\lambda-E_\mu)$,
and let us show that the granularity of the distribution of $E_\mu-E_\lambda$ in this weighted sum tends to zero faster than $\eta'/N$.
To this end, we view each eigenstate as a coherent superposition of a central phonon Fock state and of a weak halo of phonon Fock states
that emanate from the central one by Beliaev-Landau processes $1\,\mbox{phonon}\leftrightarrow 2\,\mbox{phonons}$.
Even if $\hat{W}$ preserves the phonon number, see Eq.~(\ref{eq:herisson}), it can couple the central Fock state of $|\psi_\mu\rangle$
to the halo of $|\psi_\lambda\rangle$. At worst, a single Beliaev-Landau process is involved, in which case 
$E_\mu-E_\lambda=\pm(\epsilon_{B,\kk_1}+\epsilon_{B,\kk_2}-\epsilon_{B,\kk_1+\kk_2})$,
where $\kk_1$ and $\kk_2$ are the wavevectors of the two emitted or absorbed phonons; since the pair $\{\kk_1,\kk_2\}$ takes 
a number $\propto V^2$ of different values, where $V$ is the volume of the system, $E_\mu-E_\lambda$ varies by steps $\propto 1/V^2$,
which is indeed $\ll {\gilles \eta'}/N$.
}}
\be
g_1^\lambda(t) \simeq \bar{N}_0 \eee^{-\ii{\yvan \langle \hat{W}\rangle_\lambda}t /\hbar}  \eee^{-  (\ii \delta_\lambda+ \gamma_\lambda)t  }
\ \ \ \mbox{with}\ \ \
\hbar(\delta_\lambda -\ii \gamma_\lambda) = \meanv{ \hat{W} \hat{Q}_\lambda \frac{\hat{Q}_\lambda}{(E_\lambda+\ii 0^+)\hat{Q}_\lambda - \hat{Q}_\lambda \hat{H}  \hat{Q}_\lambda } \hat{Q}_\lambda \hat{W}}_\lambda
\label{eq:carrebis}
\ee
{\yvan Remarkably, $2\gamma_\lambda$ is the decay rate of the state $|\psi_\lambda\rangle$ induced by $\hat{W}$ as predicted by Fermi's golden rule.} 

The leading term under the exponential in (\ref{eq:carrebis}) is {\yvan $\langle\hat{W}\rangle_\lambda$}, it is of order $N^0$ like $\hat{W}$. A {key} step in its interpretation is to remark that, according to the expansion in (\ref{eq:BKHpourW}), {in the Heisenberg picture
\begin{equation}
\hat{W}(t)=\hbar \frac{\dd \hat{\theta}_0}{\dd t}+O\bb{\frac{1}{N}}
\end{equation}
At this stage it may seem difficult to obtain a tractable explicit expression of ${\dd\hat{\theta}_0}/{\dd t}$ and to go beyond a purely formal result for the phase dynamics. Fortunately this is not the case and, as we will show in the next section, 
the coarse grained time average of ${\dd\hat{\theta}_0}/{\dd t}$ in a weakly excited gas is proportional to a chemical potential operator, which is in essence a thermodynamic quantity:} 
 \begin{equation}
 -\frac{\hbar}{2} \overline{ \frac{\dd\hat{\theta}_0}{\dd t} }^t=\mu_0(\hat{N})+\sum_{\substack{  s=F,B  }} \sum_{\substack{ \alpha  }} \frac{\dd \epsilon_{s,\alpha}}{\dd N} \hat{n}_{s,\alpha}
 \label{eq:herisson}
 \end{equation}
The sum on the right hand side runs over both the gapped quasiparticles fermionic branch of excitation (in the homogeneous case $\alpha$ includes both an orbital and a spin index, $\alpha=\kk,\sigma$) and the bosonic one which, in the thermodynamic limit and for an homogeneous system, has a phononic behaviour ($\alpha$ is then only orbital, $\alpha=\qq$). 
{By requiring that the gas is weakly excited we mean that the thermal depletion of the condensate of pairs must be small. This requires in particular that the average number of quasiparticles is a small fraction of the total particle number.}
The coarse grained time average is taken over a time long with respect to the inverse of the quasiparticle eigenfrequencies $\epsilon_{s,\alpha}/\hbar$ yet short with respect to the typical time-scale of variation of the {\gilles quasiparticle number operators} $\hat{n}_{s,\alpha}$,
{\gilles which is allowed if the quasiparticles are in the weakly collisional regime}. Finally $\mu_0(N)$ is the zero temperature chemical potential of the gas with $N$ particles, {that is the 
derivative of the ground state energy with respect to $N$. 
We interpret the second term on the right hand side of (\ref{eq:herisson}) as a ``chemical potential operator" in the sense that its quantum average is the adiabatic derivative of the quasiparticle gas energy $\sum_{\substack{  s=F,B  }} \sum_{\substack{ \alpha  }}  \epsilon_{s,\alpha} \langle \hat{n}_{s,\alpha} \rangle$ with respect to $N$, that is at fixed quasiparticle populations $\langle \hat{n}_{s,\alpha} \rangle$.}
{Equation (\ref{eq:herisson}) establishes the link between the phase derivative and the chemical potential at the level of quantum mechanical operators in a multimode microscopic theory. In that respect, it goes beyond the usual second Josephson relation for the phase of the superconducting order parameter (see \S3.4 in {\gilles reference} \cite{Leggett2006}).}

By taking the average of equation (\ref{eq:herisson}) in the stationary state $\ket{\psi_\lambda}$ and using the Eigenstate thermalization hypothesis \cite{Olshanii2008} to identify the quantum average in an eigenstate with the microcanonical average, we recognize the microcanonical chemical potential $\mu_{\rm mc}$ at energy $E_\lambda$ and particle number $N_\lambda$ and obtain:
\begin{equation}
\hbar {\yvan \langle \frac{\dd\hat{\theta}_0}{\dd t}\rangle_\lambda} =  -2\mu_{\rm mc}(E_\lambda,N_\lambda)
\label{eq:soleil}
\end{equation}
{We omitted here the coarse grained time average as the quantum average is taken in an exact eigenstate of the system.}

{\gilles The next term under the exponential in (\ref{eq:carrebis}) is of order $1/N$ and subleading. In order to show that,} we express this term in terms of the correlation function of  ${\dd\hat{\theta}_0}/{\dd t}$ in $\ket{\psi_\lambda}$: {up to a contribution of order $1/N^2$,}
\begin{equation}
\gamma_\lambda + \ii \delta_\lambda = \int_0^{+\infty} \dd t \bbcro{   \meanvlr{   \frac{\dd\hat{\theta}_0(t)}{\dd t} \frac{\dd\hat{\theta}_0(0)}{\dd t}  }_\lambda   -  \meanvlr{ \frac{\dd\hat{\theta}_0}{\dd t} }_\lambda^2  }   
\label{eq:corr_theta}
\end{equation}
This is equivalent to (\ref{eq:carrebis}) as can be checked by inserting a closure relation {\gilles on the eigenstates of $\hat{H}$}.
The $t=0$ value of the integrand is $\text{Var}_\lambda ({\dd\hat{\theta}_0}/{\dd t}) = O({1}/{N})$ (this comes from adding up the variances of independent quasiparticles numbers in the canonical ensemble and it overestimates the microcanonical variance);  the function then decays in a time $\tau_c$ which is the typical collision time of quasiparticles and hence the correlation time of the $\hat{n}_{s,\alpha}$. Altogether we estimate  
$|\gamma_\lambda + \ii \delta_\lambda| \approx \tau_c \text{Var} ({\dd\hat{\theta}_0}/{\dd t}) = O\bb{ {1}/{N} }$. The energy shift $\delta_\lambda$ is thus of the same order in $N$ as the {subleading} term $[\hat{\theta}_0,[\hat{\theta}_0,\hat{H}]]$ in $\hat{W}$, \textit{i.e.} $N$ times smaller than (\ref{eq:soleil}); we neglect both terms for a large system. In contrast, we keep $\gamma_\lambda$, since it is the only term which leads to an exponential decay of the correlation function $g_1^\lambda$. 
{\yvan Eq.~(\ref{eq:corr_theta}) provides a physical interpretation of $\gamma_\lambda$, 
if one remembers that, in the brownian motion theory, the integral of
the velocity correlation function gives the position diffusion coefficient.
$\gamma_\lambda$ is thus the phase diffusion coefficient of the condensate of pairs when the system is prepared in the microcanonical ensemble corresponding to $\ket{\psi_\lambda}$, hence the notation $\gamma_\lambda=D(E_\lambda,N_\lambda)$. 
We finally keep}
\begin{equation}
g_1^\lambda(t) \simeq \bar{N}_0 \eee^{2 \ii \mu_{\rm mc}(E_\lambda,N_\lambda)t/\hbar}\eee^{-D(E_\lambda,N_\lambda)t}
\label{eq:a_retenir}
\end{equation}
{\yvan Eq.~(\ref{eq:corr_theta}) even gives a way of calculating $D$: If one can write kinetic equations for the quasiparticles numbers 
appearing in $\overline{\dd\hat{\theta}_0/\dd t}^t$, see Eq.~(\ref{eq:herisson}), one can calculate their time correlation functions 
as done for bosons in reference \cite{WitkowskaCastinSinatra2009} and conclude that
\be
D(E,N)=-(P\vec{A})\cdot M^{-1} C_{\rm mc} P\vec{A} \label{eq:Dcin}
\ee
The matrix $M$, with coefficients $M_{s\alpha, s'\alpha'}$,
is the matrix of the linearised kinetic equations $\frac{\dd}{\dd t} \vec{\delta n}=M\vec{\delta n}$
that give the evolution of the fluctuations $\delta n_{s,\alpha}$ of the quasiparticle numbers, collected in a single vector $\vec{\delta n}$,
around their stationary values $\bar{n}_{s,\alpha}$.
To define the other notations, one introduces as in \cite{WitkowskaCastinSinatra2009} the covariance matrix $C_{\rm can}$ 
of the quasiparticle numbers $\hat{n}_{s,\alpha}$ in the canonical ensemble with $N$ particles and a mean energy $E$,
the energy vector $\vec{\epsilon}$ with components $\epsilon_{s,\alpha}$ and its dual vector $\vec{e} \propto C_{\rm can} \vec{\epsilon}$
normalised such that $\vec{e}\cdot\vec{\epsilon}=1.$
\footnote{{\yvan One has $(C_{\rm can})_{s\alpha,s'\alpha'}=\delta_{s\alpha,s'\alpha'} \bar{n}_{s,\alpha} (1\pm \bar{n}_{s,\alpha})$ and
$\bar{n}_{s,\alpha}=1/[\exp(\epsilon_{s,\alpha}/k_B T)\mp 1]$ where the upper (lower) sign holds for the bosonic (fermionic)
excitation branch.
$\vec{\epsilon}$ and $\vec{e}$ are left and right eigenvectors of $M$ with a zero eigenvalue,
as shown by the conservation of energy and the inspiring rewriting $\vec{e}\propto \dd\vec{\bar{n}}/\dd T$,
where $T$ is the canonical ensemble temperature and $\vec{\bar{n}}$  is the vector with components $\bar{n}_{s,\alpha}$ \cite{WitkowskaCastinSinatra2009}.}}
The vector $\vec{A}=(2/\hbar)\, \dd\vec{\epsilon}/\dd N$ then collects the coefficients of $\hat{n}_{s,\alpha}$
in the expression (\ref{eq:herisson}) of $-\overline{\dd\hat{\theta}_0/\dd t}^t$, the matrix $P$ is the non-orthogonal projector\footnote{{\yvan $P\vec{\epsilon}=\vec{0}$ and $P^\dagger$ projects onto the subspace of fluctuations $\vec{\delta n}$
of zero energy,
$\{\vec{\delta n} \,\vert\, \vec{\epsilon}\cdot\vec{\delta n}=0\}$, inside which one defines the inverse of the matrix $M$.}}
such that $P\vec{\delta n}=\vec{\delta n}-\vec{\epsilon}\, (\vec{e}\cdot \vec{\delta n})$ for all $\vec{\delta n}$,
and the microcanical covariance matrix is 
$C_{\rm mc}= P^\dagger C_{\rm can} P$ \cite{WitkowskaCastinSinatra2009}.}

The final step is to take the statistical average {\gilles of Eq.~(\ref{eq:a_retenir})} over the probability distribution $\Pi_\lambda$ of the states $\ket{\psi_\lambda}$ that constitute the mixed state of the system. For large $N$, we assume that energy and atom number fluctuations around the mean values $\bar{E}$ and $\bar{N}$ are weak in relative value. This is the case if, for example, $\Pi_\lambda$ describes a canonical or grand canonical ensemble. We assume Gaussian fluctuations and linearize $\mu_{\rm mc}$ around $(\bar{E},\bar{N})$ while, to this order, we keep only the central value $D(\bar{E},\bar{N})$ of the next-to-leading term. {We are led to the calculation of a Gaussian integral with a phase factor 
$\exp\{2\ii  [({\partial\mu_{\rm mc}(\bar{E},\bar{N})}/{\partial E})  (E-\bar{E})  +  ({\partial\mu_{\rm mc}(\bar{E},\bar{N})}/{\partial N}) (N-\bar{N})] t/\hbar\}$}. Altogether this leads to the main result of this work~:
\begin{equation}
g_1(t) \simeq \bar{N}_0 \eee^{2 \ii \mu_{\rm mc}(\bar{E},\bar{N})t/\hbar} \eee^{-{t^2}/{2t_{\rm br}^2}} \eee^{-D(\bar{E},\bar{N})t}
\label{eq:resultat}
\end{equation}
{In presence of energy or atom number fluctuations, the thermal blurring at long times} consists in a Gaussian decay of the correlation function $g_1(t)$, with a characteristic time
\begin{equation}
(2 t_{\rm br}/\hbar)^{-2}= \text{Var} \bb{ N \frac{\partial\mu_{\rm mc}}{\partial N}(\bar{E},\bar{N}) + E\frac{\partial\mu_{\rm mc}}{\partial E} (\bar{E},\bar{N}) }
\label{eq:tbr}
\end{equation}
which diverges {as} ${N}^{1/2}$ for normal fluctuations.
The phase diffusion coefficient $D$ leads to an exponential decay with a characteristic time diverging as $N$. 
As expected it is a subleading effect at long times, except if the system is prepared in the microcanonical ensemble
{in which case the intrinsic phase diffusion may be directly observed.}

\section{Microscopic derivation of the {phase operator equation}}
We give here the first (to our knowledge) microscopic derivation  of equation (\ref{eq:herisson}), relating the evolution of the phase {operator} of a pair-condensed gas to {a ``chemical potential operator"}. 

The contribution of the fermionic branch of excitations to ${\dd \hat{\theta}_0}/{\dd t}$ can be obtained from linearized equations of motion for small fluctuations of the pair operators $\hat{\psi}_\downarrow \hat{\psi}_\uparrow$, $\hat{\psi}_\uparrow^\dagger \hat{\psi}_\downarrow^\dagger$ and $\hat{\psi}_\sigma^\dagger \hat{\psi}_\sigma$ around the mean-field state {in {\gilles Anderson's} Random Phase Approximation (RPA)} \cite{Anderson1958}. Using equation (120) of reference \cite{PRA2013} to extract the time average of ${\dd\hat{\theta}_0}/{\dd t}$, and rewriting equation (86) of the same reference in terms of the fermionic quasiparticle occupation numbers {$\hat{n}_{F,\alpha}$}, 
we get\footnote{We use $\frac{\Delta_0}{\epsilon_{F,\kk,\sigma}}\hat{\zeta}_\kk=\hat{n}_{F,\kk,\uparrow}+\hat{n}_{F,-\kk,\downarrow}$ 
where {$\Delta_0$ and $\hat{\zeta}_\kk$} refer to notations of \cite{PRA2013}, and we use Eq.~(74) of that reference to recognize $\dd\epsilon_{F,\kk,\sigma}/\dd\mu$.}
\begin{equation}
-\frac{\hbar}{2}\overline{\frac{\dd\hat{\theta}_0}{\dd t}}^t\underset{\rm RPA}{=}\mu(\bar{N})+\frac{\dd\mu}{\dd\bar{N}} (\hat{N}-\bar{N}) + \sum_{\alpha=\kk,\sigma} \frac{\dd\epsilon_{F,\alpha}}{\dd\bar{N}} \hat{n}_{F,\alpha}
\label{eq:spirale}
\end{equation}
where $\epsilon_{F,\kk,\sigma}$ is the BCS excitation spectrum of an homogeneous system and $\bar{N}$ the BCS average particle number in the grand canonical ensemble of chemical potential $\mu$.
\footnote{\gilles To obtain Eq.~(\ref{eq:spirale}) in this form we reintroduced the trivial phase term $-2\mu(\bar{N})t/\hbar$ that does not appear
in the reference \cite{PRA2013} due to the use of the grand canonical Hamiltonian.}

{We encountered fundamental difficulties in deriving the phonon branch contribution to equation (\ref{eq:herisson}) within the RPA}.\footnote{{The RPA result (\ref{eq:spirale}), restricted to the linear order in the pair operators, does not include the contribution of the phonon branch.
One might hope to obtain this contribution by pushing the RPA calculation to the quadratic order in the pair operators as follows: First, one computes ${\dd \hat{\theta}_0}/{\dd t}$ up to the quadratic order. Second, one collects all the RPA pair operators inducing a center of mass momentum change $\hbar\qq$, that is $\hat{a}_{-\kk-\qq \downarrow} \hat{a}_{\kk \uparrow}$, $\hat{a}_{\kk+\qq \uparrow}^\dagger \hat{a}_{-\kk \downarrow}^\dagger$, $\hat{a}_{\kk+\qq \uparrow}^\dagger \hat{a}_{\kk \uparrow}$ and $\hat{a}_{-\kk \downarrow}^\dagger \hat{a}_{-\kk-\qq \downarrow}$, where $\hat{a}_{\kk \uparrow}$ annihilates a fermion of wave vector $\kk$ and spin $\uparrow$, and one writes the matrices ${\cal L}_\qq$ of their closed linear RPA equations of motion. Each RPA matrix ${\cal L}_\qq$ has two collective modes, with opposite energies linear in $\qq$ for small $\qq$. The RPA operators are then expanded over the eigenmodes of ${\cal L}_\qq$, with amplitudes $\hat{b}_{-\qq}$ and $\hat{b}_\qq^\dagger$ on the collective modes that annihilate and create a collective excitation of momentum $\mp \hbar \qq$.
Third, one inserts this modal expansion in the RPA operators appearing in ${\dd \hat{\theta}_0}/{\dd t}$ and obtains a quadratic expression in the modal amplitudes, hence terms in $\hat{b}_\qq^\dagger\hat{b}_\qq$ as in (\ref{eq:herisson}).
Unfortunately, this RPA approach is not reliable and must be abandoned because the RPA operators, although linearly independent, are \textit{not} quadratically independent, as one can see by rearranging the pair operators using fermionic anticommutation relations, e.g. 
\[
(\hat{a}_{\kk+\qq \uparrow}^\dagger \hat{a}_{\kk \uparrow}) (\hat{a}_{-\kk-\qq \downarrow}^\dagger \hat{a}_{-\kk \downarrow}) = (\hat{a}_{\kk+\qq \uparrow}^\dagger \hat{a}_{-\kk-\qq \downarrow}^\dagger) (\hat{a}_{-\kk \downarrow} \hat{a}_{\kk \uparrow})
\]
This shows that $(i)$ there is no unique way of expressing ${\dd \hat{\theta}_0}/{\dd t}$ as a quadratic function of the RPA operators, $(ii)$ the modal amplitudes are not quadratically independent, $(iii)$ the coefficient of $\hat{b}_\qq^\dagger\hat{b}_\qq$ is not uniquely determined by this RPA approach.}}
We therefore decided to treat the {problem variationally with  the most general  time-dependent pair coherent state Ansatz:}\footnote{{We use here for simplicity an Ansatz {\gilles of the time-dependent BCS type} in which the number of particles has quantum fluctuations (not to be confused with the thermal fluctuations of the grand canonical ensemble). The use of an Ansatz with a fixed number of particles, possible although more difficult \cite{Leggett2006}, would change 
the coefficients in the energy functional by a relative correction $O(N^{-1/2})$ and would not change the spectrum in the thermodynamic limit.}} 
\begin{equation}
\ket{\psi}= \mathcal{N}(t) \exp \bb{{l}^6 \sum_{\rr,\rr'} \Gamma(\rr,\rr';t) \hat{\psi}_\uparrow^\dagger(\rr) \hat{\psi}_\downarrow^\dagger(\rr')} \ket{0},
\label{eq:ansatz}
\end{equation}
Here $\mathcal{N}$ ensures normalization and the $\Gamma(\rr,\rr')$ form a set of independent variables.
The space has been discretized on a cubic lattice of step ${l}$,
which we take to zero in the end of the calculations. The field operators obey anticommutation relations of the kind: 
$
\{ \hat{\psi}_\sigma(\rr) , \hat{\psi}_{\sigma'}^\dagger(\rr') \}=\delta_{\sigma,\sigma'} {\delta_{\rr,\rr'}}/{{l}^3}
$.
{Section \S 9.9b of} reference \cite{Ripka1985} constructs from $\Gamma(\rr,\rr'), \Gamma^*(\rr,\rr')$ the set of canonically conjugate variables $\Phi(\rr,\rr'), \Phi^*(\rr,\rr')$.
\footnote{{\gilles Our variables $\Gamma$ and $\Phi$ correspond to the variables $z$ and $\beta$ of reference
\cite{Ripka1985} when one replaces the quasiparticle operators and vacuum in Eq.~(9.132) of \cite{Ripka1985} by the particle ones.}
If $\uuline{\Gamma}$ and $\uuline{\Phi}$ are the matrices of elements ${l}^3 \Gamma(\rr,\rr')$ and ${l}^3 \Phi(\rr,\rr')$ 
 respectively, then 
 $
 \uuline{\Phi}=-\uuline{\Gamma}(1+\uuline{\Gamma}^\dagger \uuline{\Gamma})^{-1/2}
 \label{eq:defPhi}
 $ {\gilles according to Eq.~(9.144) of \cite{Ripka1985}}. {Their variational equations of motion follow from the usual Lagrangian $L=\ii\hbar 
 \left[ \langle \psi | (\dd/{\dd}t) | \psi \rangle - \mbox{c.c.} \right]/2 -  \langle \psi | H | \psi \rangle$. $\Phi$ is 
cleverly defined such that $L=\ii\hbar l^6 \sum_{\rr,\rr'} \left[ \Phi^*(\rr,\rr') \partial_t \Phi(\rr,\rr') -\mbox{c.c.}
\right]/2-\mathcal{H}(\Phi,\Phi^*)$, leading to $\partial_t \Phi=(\ii\hbar l^6)^{-1}\partial_{\Phi^*}\mathcal{H}$.
This shows that the conjugate variable of $\Phi$ in the Hamiltonian formalism is $\ii\hbar l^6 \Phi^*$
for the usual Poisson brackets, that is $\Phi^*$ for the Poisson brackets $\{\Phi,\Phi^*\}=(\ii\hbar l^6)^{-1}$, knowing that
{\gilles one must have}
$\partial_t \Phi= \{\Phi,\mathcal{H}\}$.}}
\setcounter{noteansatz}{\thefootnote}
This field $\Phi$ {should not be confused} with the usual pairing field $\meanv{ \hat{\psi}_\downarrow \hat{\psi}_\uparrow }$.{\footnote{
{With the same matrix notation as in note \thenoteansatz\, one has $\uuline{\meanv{ \hat{\psi}_\uparrow \hat{\psi}_\downarrow }}=
\uuline{\Phi} (1-\uuline{\Phi}^\dagger \uuline{\Phi})^{1/2}$, {\gilles according to Eq.~(9.146) of the reference \cite{Ripka1985}}.}}}
\setcounter{notepaire}{\thefootnote}
 {When the pairs in (\ref{eq:ansatz}) are at rest, $\Gamma(\rr,\rr',t)$ depends only on $\rr-\rr'$ and the Fourier transform of $\Phi$ reproduce the $V_{\kk}$ amplitude of the $\kk\uparrow$, $-\kk\downarrow$ pair of BCS theory \cite{PRA2013}, while the Fourier transform of the pairing field is 
$-U_\kk V_\kk=-V_\kk(1-|V_\kk|^2)^{1/2}$. For moving pairs we have no physical interpretation, but the squared norm of $\Phi$ is still 
half the mean number of particles in the state $\ket{\psi}$:}
\begin{equation}
\frac{N}{2}=\Vert \Phi \Vert^2 \equiv {l}^6 \sum_{\rr,\rr'}  |\Phi(\rr,\rr';t)|^2.
\label{eq:nuage}
\end{equation}
{\gilles The classical Hamiltonian governing the evolution of the field $\Phi$ is given by}
\begin{equation}
\mathcal{H}(\Phi,\Phi^*)=\bra{\psi} \hat{H} \ket{\psi} .
\end{equation}
{\gilles and may be expressed explicitly using Wick's theorem.}
In the following however, we will only need the invariance of $\mathcal{H}$ under a global phase change 
 $\Phi(\rr,\rr') \rightarrow \eee^{i\gamma}\Phi(\rr,\rr')$, $\forall \gamma \in \mathbb{R}$ ($U(1)$ symmetry), 
consequence of the conservation of the particle number $\hat{N}$ by evolution with $\hat{H}$. 
At zero temperature and for a fixed $N$ the field $\Phi(\rr,\rr')$ is frozen, up to a global phase factor, into the minimizer $\Phi_0(\rr,\rr')=(N/2)^{1/2} \phi_0(\rr,\rr')$ of $\mathcal{H}$. $\phi_0$ is chosen real and normalized to one. It depends on $N$ 
even in the spatially homogeneous case and differs from the condensate wavefunction  $\varphi_0$ in the same way that 
 $\Phi$ differs from the pairing field $\meanv{ \hat{\psi}_\downarrow \hat{\psi}_\uparrow }$ {(see note \thenotepaire)}.
At sufficiently low temperature one can expand $\mathcal{H}$  in powers of the small deviations of $\Phi$ away from
the circle $\gamma \mapsto \eee^{i\gamma} \Phi_0(\rr,\rr')$, locus of  the minima of $\mathcal{H}$ for fixed $N$.
{To this end,} we split the field into its components parallel and orthogonal to $\phi_0$ :
\begin{equation}
\Phi(\rr,\rr')=\eee^{i\theta}[n^{1/2}\phi_0(\rr,\rr')+\Lambda(\rr,\rr')]
\label{eq:fleur}
\end{equation}
The phase $\theta$ can reach arbitrarily large values while $\Lambda$ is bounded. This framework allows us to develop
a systematic perturbation theory in powers of the field $\Lambda$ (see appendix A), {the} fermionic equivalent of the Bogoliubov $U(1)$-symmetry conserving approach for bosons \cite{CastinDum1998}.
Provided that $\Lambda$ stays small, the phase $\theta$ remains close to the condensate phase $\theta_0$ as we shall see.
We therefore write down the equations of motion of  $\theta$ and of the fields $\Lambda, \Lambda^*$. 
At the end of the calculations we systematically eliminate the condensate variables with the relation  $n={\gilles N/2} - \Vert \Lambda \Vert^2$, consequence of (\ref{eq:nuage}), and we restrict ourselves to order 2 in $\Lambda, \Lambda^*$. 

The main challenge of the calculation is the occurrence of a term linear in $\Lambda,\Lambda^*$ in ${\dd \theta}/{\dd t}$, resulting from the fact that $\phi_0$ depends on the number of pairs \cite{PRA2013}. Without this term,
one would simply expand the field $\Lambda$ on the eigenmodes of its small linear oscillations obtained
from a quadratization of the Hamiltonian $\mathcal{H}$ at fixed $N$:
\begin{equation}
\begin{pmatrix}
\Lambda(\rr,\rr';t) \\ 
\Lambda^*(\rr,\rr';t)
\end{pmatrix}
=
\sum_\alpha b_\alpha(t) 
\begin{pmatrix}
u_\alpha(\rr,\rr') \\ 
v_\alpha(\rr,\rr')
\end{pmatrix} 
+ b_\alpha^*(t)
\begin{pmatrix}
v_\alpha^*(\rr,\rr') \\
u_\alpha^*(\rr,\rr')  
\end{pmatrix} 
\label{eq:doubleetoile}
\end{equation}
where the sum runs over the eigenmodes of positive energy $\epsilon_\alpha$, 
normalized as $\Vert u_\alpha \Vert^2 - \Vert v_\alpha \Vert^2 =1$. To this order, 
$b_\alpha(t) = b_\alpha(0) \eee^{-i\epsilon_\alpha t/\hbar}$.
One would insert the expansion (\ref{eq:doubleetoile}) into ${\dd \theta}/{\dd t}$ and take a coarse grained temporal average to get rid of the oscillating terms. $\overline{{\dd \theta}/{\dd t}}^t$ would then contain the expected linear
combination of the numbers of bosonic quasiparticles $n_{B,\alpha}=|b_\alpha|^2$.
In reality, the problem is more subtle: due to the interaction among the quasiparticles, $\overline{b_\alpha}^t$ does not vanish and is of order two in $\Lambda$ and $\Lambda^*$. The contribution of the linear term in ${\dd \theta}/{\dd t}$ is then comparable to that of the quadratic ones. It is calculated in the appendix A, using in particular
the bounded nature of the field $\Lambda$ (a consequence of the $U(1)$-symmetry preserving nature of the expansion (\ref{eq:fleur}))
and the Hellmann-Feynman theorem. One finds
\begin{equation}
-\frac{\hbar}{2} \overline{\frac{\dd \theta}{\dd t}}^t =\mu_0(N)+\sum_\alpha \frac{\dd \epsilon_\alpha}{\dd N} |b_\alpha|^2+O(\Vert \Lambda \Vert^3)
\label{eq:bla}
\end{equation}

We now briefly discuss the form of the energy spectrum $\epsilon_\alpha$ for a spatially homogeneous system, in the continuous limit ${l}\rightarrow 0$
for a $s$-wave contact interaction with a fixed scattering length between opposite spin fermions. For each value of the total wave vector $\qq$, there exists $(i)$ at most one discrete
value $\epsilon_{B,\qq}$, $(ii)$ a continuum parametrized by two wave vectors $(\kk_1,\uparrow;\kk_2,\downarrow) \mapsto \epsilon_{F,\kk_1,\uparrow} + \epsilon_{F,\kk_2,\downarrow}$ of constant sum ($\kk_1+\kk_2=\qq$), where $\epsilon_{F,\kk,\sigma}$ is the BCS dispersion relation.
The branch $\epsilon_{B,\qq}$ coincides with that of reference \cite{CKS2006}, as we have checked. It has a phononic start and corresponds to the bosonic elementary excitations of the Fermi gas, whose contribution to the phase dynamics was missing. 
The continuum corresponds to the excitation of two fermionic quasiparticles. Indeed, since the Hamiltonian $\hat{H}$ contains an even number of factors $\hat{\psi}$ and
$\hat{\psi}^\dagger$, each annihilating or creating one quasiparticle, fermionic  quasiparticles can
only be created by pairs from the ground state. The corresponding biexcitations are not physically independent,\footnote{Exciting $\alpha=(\kk_1,\uparrow;\kk_2,\downarrow)$ and $\alpha'=(\kk_1',\uparrow;\kk_2',\downarrow)$ amounts to exciting $\alpha''=(\kk_1',\uparrow;\kk_2,\downarrow)$ and $\alpha'''=(\kk_1,\uparrow;\kk_2',\downarrow)$.} 
and {\yvan duplicate the contribution of the RPA to ${\dd \hat{\theta}_0}/{\dd t}$. They must therefore not be included in Eq.~(\ref{eq:bla}).}

Two more remarks are needed to obtain (\ref{eq:herisson}). $(i)$ The fields $\meanv{\hat{\psi}_\downarrow\hat{\psi}_\uparrow}$ and $\Phi$ differ and so do the phases $\hat{\theta}_0$ and $\theta$. Their coarse grained temporal averages, however, only differ by a term of order $\Vert \Lambda \Vert^2$, which, bounded hence negligible in the long time limit, does not contribute to the phase blurring of the condensate of pairs.{\footnote{{Expressing in $\meanv{\hat{a}_0}$ the pairing field in terms of $\Phi$, one realizes that, for small $\Lambda$, $\meanv{\hat{a}_0}=e^{i\theta} \sqrt{N_0} (1+O(||\Lambda||))$ and, since $\overline{\Lambda}^t=O(||\Lambda||^2)$, one has $\overline{\theta_0}^t=\overline{\theta}^t+O(||\Lambda||^2)$.}}} $(ii)$ The phase $\theta$ of our variational approach is a classical variable, whereas $\hat{\theta}_0$ in (\ref{eq:herisson}) is a quantum operator. This gap can be bridged by using the quantization procedure exposed in Chapter 11 of reference \cite{Ripka1985} 
where the $b_\alpha$ {of the bosonic branch}
are in the end replaced by bosonic operators{\footnote{{
More precisely, these operators are bosonic only for a weak density of excitations.
For a spatially homogeneous system and in a $U(1)$ symmetry breaking formalism ($\theta=0$), 
we obtain from Eq.(11.81c) of \cite{Ripka1985} extended to the paired case ($\hat{a}_p,\hat{a}_h^\dagger$ 
replaced by the BCS fermionic quasiparticle annihilation operators $\hat{b}_{\kk\sigma}$) and restricted to weakly excited bosonic 
images ($\mathbb{B}\mathbb{B}^\dagger$ negligible):
$\hat{b}_\qq = \sum_{\kk} X_\kk^\qq \hat{b}_{-\kk\downarrow} \hat{b}_{\kk+\qq\uparrow} + 
Y_\kk^\qq \hat{b}_{\kk-\qq\uparrow}^\dagger \hat{b}_{-\kk\downarrow} ^\dagger$.
The real coefficients $X^\qq$ and $Y^\qq$ are linear combinations of the corresponding $u_\qq$ and $v_\qq$
in Eq.(\ref{eq:doubleetoile}), and inherit the normalization
condition $\sum_\kk (X_\kk^\qq)^2-(Y_\kk^\qq)^2=1$. 
Then in a state (\ref{eq:ansatz})
with $\theta=0$, $\hat{\delta}_\qq\equiv [\hat{b}_\qq,\hat{b}_\qq^\dagger]-\hat{1}$ has a mean value and a variance
$O(||\Lambda||^2/N)$ since $\hat{b}_{\kk\sigma} |\psi\rangle=O(||\Lambda||/N^{1/2})$ if $||\Lambda||/N^{1/2}\!\to\!0$.
   }}} 
$\hat{b}_\alpha$, $[\hat{b}_\alpha,\hat{b}_\alpha^\dagger]=1$.
We argue that equation (\ref{eq:herisson}), linking $\dd \hat{\theta}_0 / \dd t$ to the chemical potential operator, and the resulting equation (\ref{eq:resultat}) should hold beyond the validity range of the microscopic variational derivation presented above, and should apply
even to the strongly interacting regime, provided that the temperature is low enough for the quasiparticles lifetime to be much longer than the inverse of their eigenfrequency.
Indeed, in the limiting case where one can neglect the fermionic excitation branch and drop the non-phononic part of the bosonic branch, Eq.~(\ref{eq:herisson}) can be derived from the irrotational version of the quantum hydrodynamic theory of Landau and Khalatnikov \cite{Khalatnikov1949} (see appendix B).

\section{Towards an experimental observation}
\begin{figure}[t]
\centerline{\includegraphics[width=5cm,height=2.8cm,clip=]{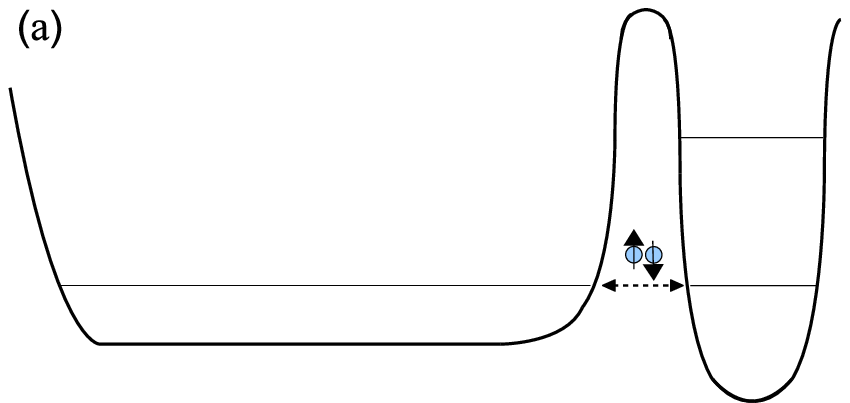} \hspace{2cm}\includegraphics[width=6.5cm]{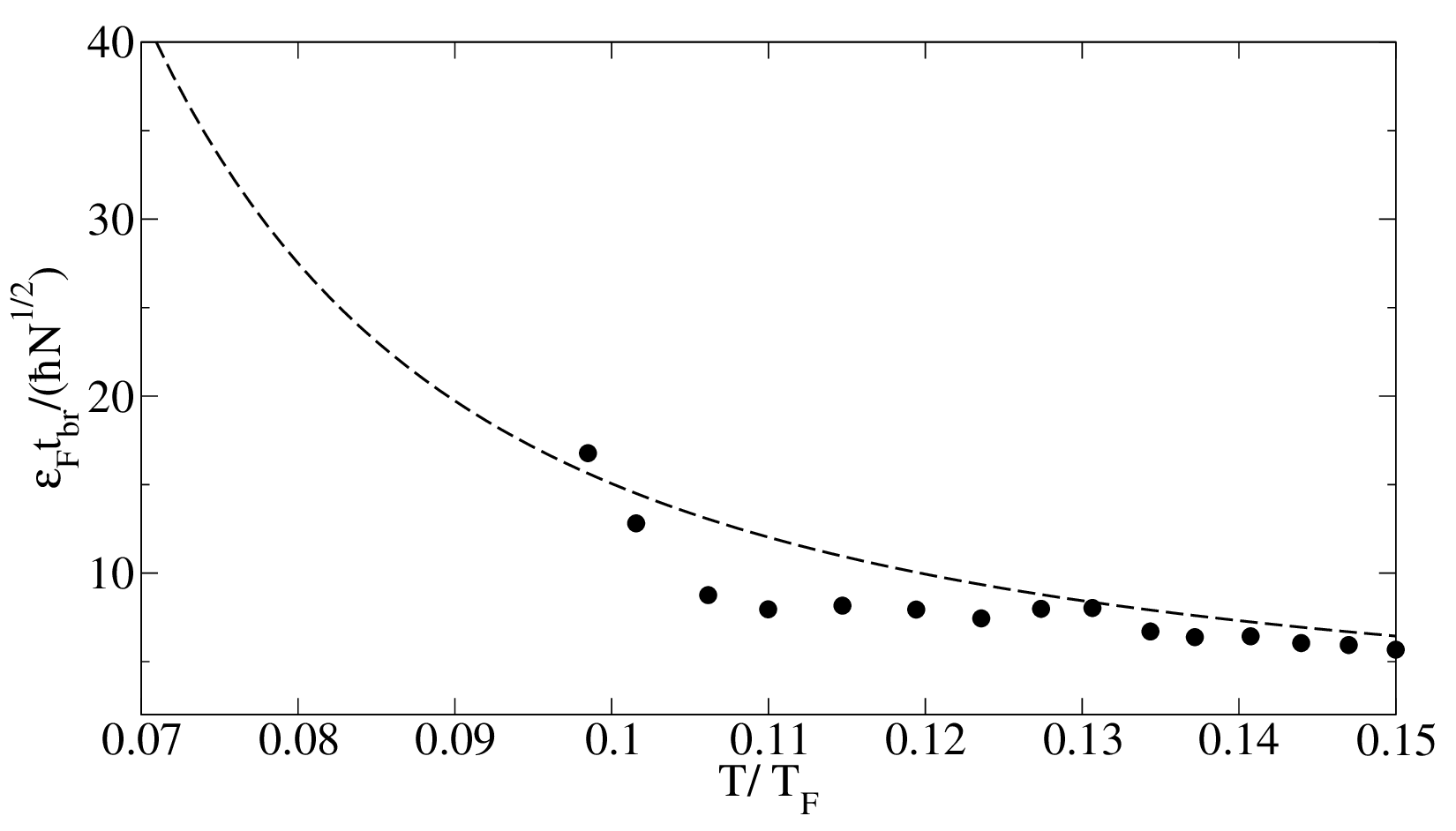}}
\caption{Some experimental considerations: (a) Trapping configuration proposed here to measure $g_1(t)$ via Ramsey interferometry: the condensed paired Fermi gas is confined in the main trap (with a flat bottom on the figure); one transfers on average at most two atoms (in the form of a dimer) in the (very narrow) secondary trap via a resonant tunneling effect, which can be tuned by a barrier of adjustable height; in this way, one creates a phase reference, which is made to interfere with the condensate after an evolution time~$t$.
(b) Thermal blurring time of a coherent Fermi gas in the unitary limit in the canonical ensemble, as a function of temperature
$T$ in units of the Fermi temperature $T_F=\varepsilon_F/k_B$. Discs: from the equation of state measured in 
reference \cite{Zwierlein2012}.  Dashed line: expression (\ref{eq:3jours}) deduced from an approximated equation of state (see text).
\label{fig:piege_tbr}}
\end{figure}

Let us briefly explain how an experimental evidence of the thermal blurring of a condensate of pairs could be obtained. The key idea is to bosonize the atomic Cooper pairs into deeply bound weakly interacting dimers during the preparation and the measurement stage. This can be done in an adiabatic reversible way \cite{Grimm2004} by tuning the scattering length to a small and positive value thanks to a magnetic Feshbach resonance. It allows one to $(i)$ produce a sample of dimers with weak number fluctuations from a melted Mott phase of an experimental realization of the Bose Hubbard model \cite{Bloch2002}, 
$(ii)$ control tunneling between the main trap (containing the $N$ particles) and a very narrow secondary trap by adjusting the height of a potential barrier \cite{Oberthaler2008} (Fig.\ref{fig:piege_tbr}a), 
$(iii)$ detect by fluorescence a single dimer \cite{Bakr2009} in this secondary trap. For the measurement of the $g_1(t)$ function,
we adapt to the case of paired fermions the interferometric Ramsey method of reference \cite{CastinSinatra2012}, where two Rabi pulses are applied at a time interval $t$. The bosonized pairs are prepared initially in the main trap. A first pulse of angle $\epsilon$ transfers on average less than one dimer to the secondary trap; in this way, the thermal blurring is not masked by partition noise. Then the system evolves during a time $t$ with interactions set to the value at which phase dynamics is to be studied. Last, the gas is rebosonized and a second pulse of angle $\epsilon$ closes the interferometer before the number $n_{\rm sec}$ of dimers in the secondary trap is measured.
The average of $n_{\rm sec}$ over the realizations is an oscillating function of the time $t$, of angular frequency $2/\hbar$ times
the difference of the two trapping zones chemical potentials, and of contrast equal to $|g_1(t)/g_1(0)|$.

\section{{Explicit results for the unitary gas}}
{We first} estimate the blurring time for a unitary Fermi gas prepared in the canonical ensemble, that is with energy fluctuations of variance $\mbox{Var}\,E=k_BT^2 \partial_T \bar{E}$. From the equation of state of the unpolarized unitary gas measured in
reference \cite{Zwierlein2012}, and for a spatially homogeneous system (in a flat bottom potential \cite{Hadzibabic2013}) we find the thermal blurring times $t_{\rm br}$ plotted as discs in figure \ref{fig:piege_tbr}b.
For example, at a temperature $T=0.12\ T_F\simeq 0.7\ T_c$, {\gilles where $T_c$ is the pair condensation temperature},
 we find $t_{\rm br}\approx 7 N^{1/2}\hbar/\varepsilon_F$ corresponding
to $20$ milliseconds for a typical Fermi temperature $T_F=\varepsilon_F/k_B=1 \mu \text{K}$ and a typical atom number $N=10^5$.
As in reference \cite{Bulgac2006}, one can also estimate the equation of state of the unitary gas from simple dispersion relations for the elementary
excitations. For the bosonic branch one takes \cite{Bulgac2006} $\epsilon_{B,\qq}=\hbar c q$ with $c$ the $T=0$ sound velocity,
$m c^2=\frac{2}{3} \xi \varepsilon_F$ and $\xi$ the Bertsch parameter. For the fermionic branch, one takes \cite{Son2006} $\epsilon_{F,\kk,\sigma}=\Delta+(\frac{\hbar^2 k^2}{2m}-\varepsilon_0)^2/(2{f_0})$, where $\Delta$ is the gap, and $\varepsilon_0$ and ${f_0}$ give the location of the minimum and the curvature of the dispersion relation. Keeping each branch contribution {\gilles to the mean volumic energy} to its leading order at low temperature {\gilles $\bar{E}_F/V\simeq \frac{2m^{3/2}}{\pi^{3/2}\hbar^3} (\epsilon_0 f_0 k_B T)^{1/2} \Delta \eee^{-\Delta/k_B T}$ and
$\bar{E}_B/V\simeq \frac{\pi^2}{30} (k_B T)^4/(\hbar c)^3$} as in \cite{Bulgac2006} and using the experimental values \cite{Zwierlein2012,Ketterle2008} $\xi=0.376$, $\Delta=0.44\varepsilon_F$, $\varepsilon_0=0.85\varepsilon_F$
and the theoretical value \cite{Son2006} ${f_0}=0.846\varepsilon_F$, we find 
\begin{equation}
\frac{N\hbar^2}{(t_{\rm br} \varepsilon_F)^2 }\simeq \bb{\frac{\Theta}{{0.46}}}^5{\frac{(1+2r)^2}{(1+r)}} 
\label{eq:3jours}
\end{equation}
{with $\Theta=T/T_F$} and $r{\gilles\equiv\partial_T \bar{E}_F/\partial_T \bar{E}_B}\simeq\bb{\frac{0.316}{\Theta}}^{9/2} \eee^{-0.44/\Theta}$ the relative weight of the two excitation branches.
This formula,\footnote{The good agreement with the experimental data has to be taken cautiously. If one treats the two branches to all order in $k_B T$, one gets an upward shift of $t_{\rm br} \varepsilon_F /(\hbar N^{1/2})$ more or less constant and equal to 5 over the temperature range of figure \ref{fig:piege_tbr}b.} plotted as a dashed line in figure \ref{fig:piege_tbr}b, is an exact equivalent to $t_{\rm br}$ {for} $\Theta\rightarrow0$.

At finite times\footnote{{\gilles Still the times that we consider are subjected to the constraint  $t>\tau_c$ (see note \thenotetauc). For the unitary gas $\tau_c{\simeq} (0.149/\Theta)^{5}$
when $\Theta \rightarrow 0$, according to note \ref{note:D}.}} 
\setcounter{notetaucexpli}{\thefootnote}
$t=O(N^0)\ll t_{\rm br}$, the contribution of $D$ to $g_1(t)$ in equation (\ref{eq:resultat}) is \textit{a priori} comparable to that of 
$t_{\rm br}^{-2}$ since both scale as $1/N$ in the canonical ensemble. $D$ can be calculated at very low temperatures keeping only the phononic part of the spectrum. 
{\yvan In the state of the art, it is predicted by various approaches that the bosonic branch is convex at low $q$ for the unitary gas
\cite{Castin2015,Salasnich2015,Tempere2011},
\be
{\yvan
\epsilon_{B,\qq}\underset{q\to 0}{=} \hbar c q + \frac{\gamma \hbar^3 q^3}{8 m^2 c} + O(q^5)\ \ \mbox{with}\ \ \gamma>0
\label{eq:def_courbure}
}
\ee
so that phonons interact through Landau-Beliaev processes {\yvan $2\,\mbox{phonons} \leftrightarrow 1\,\mbox{phonon}$} 
as in the weakly interacting Bose gas.\footnote{{\yvan
In the concave case, the leading processes are the scattering events $2\,\mbox{phonons}\leftrightarrow 2\,\mbox{phonons}$ as in
reference \cite{Khalatnikov1949}.}} 
One can then adapt
\footnote{{\yvan
\label{note:D}
The universalising trick at low temperature is to use, whatever the physical system, $m c/\hbar$  as the unit of wavevector and $m c^2$
as the unit of energy. Let us then start from the phase diffusion coefficient and the quasiparticle correlation time for
weakly interacting bosons \cite{WitkowskaCastinSinatra2009},
$\hbar N D/(m c^2)\sim c_1 (k_B T/m c^2)^4$ and $ mc^2 \tau_c/\hbar \sim
c_3 \rho [\hbar/(\sqrt{2} mc)]^3 (m c^2/k_B T)^5$,
where $c_1\simeq0.3036$ and $c_3\simeq 0.05472$, and let us review all the corrective factors connecting the weakly interacting
Bose gas to the unitary Fermi gas, knowing that $D$ is a quadratic function of $P\vec{A}$ and a linear function of $M^{-1}$,
see Eq.~(\ref{eq:Dcin}): 
$(i)$ for a condensate of pairs, there is an additional factor two in the coefficients $\vec{A}$ of
$-\overline{\mathrm{d}\hat{\theta}_0/\mathrm{d}t}^t$, thus a factor four on $D$,
$(ii)$ the equation of state $\mu_0(\rho)\propto \rho$ is replaced by $\mu_0(\rho)\propto \rho^{2/3}$,  where $\rho$ is the density,
so the value of $(N/c)\mathrm{d}c/\mathrm{d}N$ in $\vec{A}$ jumps from $1/2$ to $1/3$, hence a factor $2/3$ on $\vec{A}$
and a factor $4/9$ on $D$,
$(iii)$ $P\vec{A}$, being exactly zero for a linear-in-$q$ bosonic excitation branch due to $\vec{A}\propto \vec{\epsilon}$,
is proportional at low temperature to the dimensionless curvature parameter $\gamma$, equal to one in \cite{WitkowskaCastinSinatra2009}, 
hence an extra factor $\gamma^2$ in $D$, $(iv)$ according to quantum hydrodynamics \cite{Khalatnikov1949}, the reduced Beliaev-Landau coupling
applitues among the quasiparticles, at fixed values of the reduced wavevectors,
are system-dependent only {\sl via} a global factor $1+\Lambda$, with $\Lambda=\rho \mu_0''(\rho)/[3\mu_0'(\rho)]$ jumping from
$0$ to $\Lambda_u=-1/9$, hence a factor $(1+\Lambda_u)^{-2}$ on the matrix $M^{-1}$, on $D$ and on the correlation time $\tau_c$ 
induced by the collisions among the quasiparticles. As a consequence, for the unitary gas,
$\hbar D N/(mc^2) \sim c_1 [(4\gamma/3)^2/(1+\Lambda_u)^2] (k_B T/m c^2)^4$
and $mc^2 \tau_c/\hbar \sim c_3 (1+\Lambda_u)^{-2} \rho [\hbar/(\sqrt{2} mc)]^3 (m c^2/k_B T)^5$.
}}
the low-temperature reduction of the general expression (\ref{eq:Dcin}) done in reference \cite{WitkowskaCastinSinatra2009}.
Furthermore, from a numerical solution of the RPA equations of reference \cite{CKS2006} we find that $\gamma\approx 0.1$, in agreement
with \cite{Tempere2011,KCS2016}, so that}
\begin{equation}
\frac{\hbar N D}{ \varepsilon_F} \underset{\Theta \to 0}{\sim} C \, \Theta^4 \, \qquad  \mbox{with}\qquad
C\simeq 0.4
\label{eq:D_unitary}
\end{equation}
With this we reach a complete picture of the thermal blurring of the unitary Fermi gas at low temperature. 

\section{Conclusion}
We have presented the first microscopic theory of the thermal blurring of the phase of a condensate of pairs of fermions (\ref{eq:resultat}), revealing a ballistic blurring and a subleading phase diffusion. The blurring time depends on the variance of the total energy of the gas, and on the derivative of the microcanonical chemical potential with respect to the energy. 
{\gilles The phase diffusion coefficient can be deduced from kinetic equations on the quasiparticle numbers.
To obtain these results, we have used the fact that}
the time derivative of the condensate phase is given by the chemical potential operator of the gas, see equation (\ref{eq:herisson}). We have derived this central relation in a fully microscopic way, including both the bosonic and the fermionic branches of excitation. Last, we have proposed a realistic experimental protocol to measure this blurring time,
that we estimated to be tens of milliseconds for a coherent gas prepared in the unitary limit in the canonical ensemble.


\smallskip
\section*{Acknowledgments}
We acknowledge support from the EU project QIBEC under contract number 284 584. 




\section*{Appendix A.\ \ More on the variational calculation}

{Here we derive equation (\ref{eq:bla}) of the time average of $\theta$ within the microscopic model based on the Ansatz (\ref{eq:ansatz}), coherent state of moving pairs.} In a first stage one should  perform the expansion of the Hamiltonian $\mathcal{H}$ treating the real quantity $n$ and the complex field $\Lambda$ {\gilles in (\ref{eq:fleur})} as
\textit{independent} variables, that is, not fixing the value of $\Vert \Phi \Vert$. To include interactions
among the quasiparticles, one must go to third order in $\Lambda$ and $\Lambda^*$:
\begin{equation}
\mathcal{H}(\Phi,\Phi^*)=\mathcal{T}_0[n,\phi_0(N)]+\sum_{j=1}^3 \mathcal{T}_j[n,\phi_0(N)](\Lambda,\Lambda^*) + O(\Vert \Lambda \Vert^4)
\label{eq:carre}
\end{equation}
where the tensor $\mathcal{T}_j$ is of rank $j$ so that $\mathcal{T}_j(\Lambda,\Lambda^*)$ is exactly of order
$j$ in $\Lambda$ and $\Lambda^*$.  It may be expressed in terms of the differential of order
$j$ of $\mathcal{H}$ taken at $(\Phi,\Phi^*)=(n^{1/2}\phi_0,n^{1/2}\phi_0)$ and restricted to the subspace
orthogonal to $(\phi_0,0)$ and $(0,\phi_0)$, {\gilles with $\phi_0$ assumed to be real}. It does not depend on the phase $\theta$ due to 
$U(1)$ symmetry. For a fixed total number of particles, the energy does not vary to first order 
around the minimizer
so that $\mathcal{T}_1[N/2,\phi_0(N)]=0$, which is the famous
gap equation when the system is spatially homogeneous. Furthermore, one can check 
that $\partial_n \mathcal{T}_0[N/2,\phi_0(N)]=2\mu_0(N)$ where $\mu_0(N)=\dd E_0(N)/\dd N$ is the gas
chemical potential at zero temperature, $E_0(N)=\mathcal{T}_0[N/2,\phi_0(N)]$ being the ground state energy.

The phase and the modulus square of the {component} of the field $\Phi$ on the mode $\phi_0$ are canonically
conjugate variables, so that $-\hbar{\dd \theta}/{\dd t}=\partial_n \mathcal{H}(\Phi,\Phi^*)$.
Once this derivative is taken in (\ref{eq:carre}) for fixed $\Lambda$ and $\Lambda^*$,
one can fix the norm of $\Phi$ to the value $(N/2)^{1/2}$ (that is the total particle number is fixed to $N$),
and eliminate $n$ through the identity $n=\Vert \Phi \Vert^2 - \Vert \Lambda 
\Vert^2$; the field $\Lambda$ then remains the only dynamical variable of the problem.
The resulting expression is useful up to order 2 in $\Lambda,\Lambda^*$:
\be
-\hbar\frac{\dd \theta}{\dd t}=\partial_n \mathcal{T}_0[{N}/{2},\phi_0(N)]-\Vert \Lambda \Vert^2 \partial_n^2 \mathcal{T}_0[{N}/{2},\phi_0(N)] + \sum_{j=1}^2 \partial_n \mathcal{T}_j[{N}/{2},\phi_0(N)](\Lambda,\Lambda^*) + O(\Vert \Lambda \Vert^3)
\label{eq:triangle}
\ee

The Hamiltonian that determines the evolution of $\Lambda$ at fixed particle number is obtained by replacing
$n$ with $N/2-\Vert \Lambda \Vert^2$ in (\ref{eq:carre}) and by expanding the resulting expression up to order three in $\Lambda, \Lambda^*$:
\be
\mathcal{H}_N(\Lambda,\Lambda^*)=E_0(N)+ \check{\mathcal{T}}_2[N](\Lambda,\Lambda^*) + {\mathcal{T}}_3[N/2,\phi_0(N)](\Lambda,\Lambda^*) 
-\Vert \Lambda \Vert^2 \partial_n \mathcal{T}_1[N/2,\phi_0(N)](\Lambda,\Lambda^*) + O(\Vert \Lambda \Vert^4)
\label{eq:sapin}
\ee
with the quadratic form $\check{\mathcal{T}}_2[N](\Lambda,\Lambda^*)$ obtained by 
subtracting $2\mu_0(N)\Vert \Lambda \Vert^2$ from ${\mathcal{T}}_2[N/2,\phi_0(N)](\Lambda,\Lambda^*)$.
{To compute the coarse grained time average of $\partial_n \mathcal{T}_1[{N}/{2},\phi_0(N)](\Lambda,\Lambda^*)$,} 
{\gilles that is the term linear in $\Lambda, \Lambda^*$ which is problematic in $\dd\theta/\dd t$ (see the main text of the paper),}
we write the temporal derivative of the imaginary part of the component of
the field $\Lambda$ on the function $(N/2)^{1/2} \dd \phi_0/\dd N$,
\begin{equation}
Y{\gilles \equiv}\frac{{l}^6}{2\ii} \sum_{\rr,\rr'} \bb{\frac{N}{2}}^{1/2} \frac{\dd \phi_0(\rr,\rr')}{\dd N} (\Lambda(\rr,\rr')-\Lambda^*(\rr,\rr'))
\end{equation}
Since $\ii \hbar \partial_t \Lambda={l}^{-6} \partial_{\Lambda^*} \mathcal{H}_N(\Lambda,\Lambda^*) $, one gets
\be
-2\hbar \frac{\dd Y}{\dd t}= {\mathcal{D}} \cdot \mathcal{H}_N(\Lambda,\Lambda^*)={\mathcal{D}}\cdot\check{\mathcal{T}}_2[N](\Lambda,\Lambda^*)
+{\mathcal{D}}\cdot\mathcal{H}_N^{\rm cub}(\Lambda,\Lambda^*)+O(\Vert \Lambda \Vert^3)
\label{eq:spirale2}
\ee
where $\mathcal{H}_N^{\rm cub}$ is the component of $\mathcal{H}_N$ of order three in $\Lambda,\Lambda^*$. 
We have introduced the differential operator
\begin{equation}
{\mathcal{D}}= \sum_{\rr,\rr'} \bb{\frac{N}{2}}^{1/2} \frac{\dd \phi_0(\rr,\rr')}{\dd N} \bb{\partial_{\Lambda(\rr,\rr')}+\partial_{\Lambda^*(\rr,\rr')}}
\label{eq:operateur_differentiel}
\end{equation}
We shall now take advantage of two identities that exactly hold for all $\Lambda$ orthogonal to $\phi_0$:
\begin{eqnarray}
2{\mathcal{D}}\cdot \check{\mathcal{T}}_2[N](\Lambda,\Lambda^*)&\!\!\!\!\!\!=\!\!\!\!\!\!&-\partial_n \mathcal{T}_1{\gilles [N/2,\phi_0(N)]} (\Lambda,\Lambda^*) \label{eq:cascade1} \\
2{\mathcal{D}}\cdot\mathcal{H}_N^{\rm cub}(\Lambda,\Lambda^*)&\!\!\!\!\!\!=\!\!\!\!\!\!& 2\frac{\dd}{\dd N}\check{\mathcal{T}}_2{\gilles [N]}(\Lambda,\Lambda^*) - \partial_n \mathcal{T}_2{\gilles [N/2,\phi_0(N)]}(\Lambda,\Lambda^*) + \Vert \Lambda \Vert^2 \partial_n^2 \mathcal{T}_0{\gilles [N/2,\phi_0(N)]}  \label{eq:cascade2}  
\end{eqnarray}
To prove these relations, one formally considers a field $\Phi'$ with $N+\delta N$ particles
and determines in two different ways the quadratic expansion of $\mathcal{H}_{N+\delta N}(\Lambda',\Lambda'^*)$
in powers of $\Lambda'$ and $\Lambda'^*$, where $\Lambda'$ is {\gilles up to a global phase}
as in (\ref{eq:fleur}) (written for $N+\delta N$ particles) the component of $\Phi'$
orthogonal to $\phi_0(N+\delta N)$. First, one simply replaces $N$ with $N+\delta N$ and $\Lambda$
with $\Lambda'$ in (\ref{eq:sapin}), and then expands to first order in $\delta N$. The tensor $\dd \check{\mathcal{T}}_2/\dd N [N]$ naturally appears from this expansion. Second, one applies to $\mathcal{H}(\Phi',\Phi'^*)$
the expansion (\ref{eq:carre}) around $\Phi_0(N)$ and takes into account the fact that, to first order
in $\delta N$, the component of $\Phi'$ orthogonal to $\phi_0(N)$ contains, in addition 
to $\Lambda_\perp'$ coming from $\Lambda'$, a contribution coming from $\dd \phi_0/\dd N$:
\be
\Lambda{\gilles (\rr,\rr')}=\delta N \bb{\frac{N}{2}}^{1/2} \frac{\dd \phi_0}{\dd N}{\gilles (\rr,\rr')}+\Lambda_\perp'{\gilles (\rr,\rr')}
+O(\delta N^2,||\Lambda'||^{\gilles 3},\delta N ||\Lambda'||^{\gilles 2})
\label{eq:wallaby}
\ee
This infinitesimal shift proportional to $\delta N$ along the direction of $\dd \phi_0/\dd N$ is responsible for the occurrence 
of the operator ${\mathcal{D}}$. Equations (\ref{eq:cascade1}) and (\ref{eq:cascade2}) are finally obtained by identification
of the two resulting expressions of
 $\mathcal{H}_{N+\delta N}(\Lambda',\Lambda'^*)$ respectively to first and second
order in $\Lambda'$ and $\Lambda'^*$.

It remains to combine equations (\ref{eq:triangle}), (\ref{eq:spirale2}), (\ref{eq:cascade1}),
and (\ref{eq:cascade2}) after a coarse grained temporal average 
(over a time scale much longer than the oscillation period of the
modal amplitudes $b_\alpha$ but much shorter than the evolution time of the quasiparticle numbers $|b_\alpha|^2$),
to obtain 
\begin{equation}
-\frac{\hbar}{2} \overline{\frac{\dd \theta}{\dd t}}^t = \mu_0(N)+\overline{\frac{\dd \check{\mathcal{T}}_2}{\dd N} [N] (\Lambda,\Lambda^*)}^t + O(\Vert\Lambda\Vert^3)
\label{eq:blu}
\end{equation}
where we used the crucial property that $\overline{{\dd \Lambda}/{\dd t}}^t$ vanishes (and so does $\overline{{\dd Y}/{\dd t}}^t$), since the range of variation of the field $\Lambda$ is bounded. The quadratic form $\check{\mathcal{T}}_2[N]$ is represented
by the matrix $\eta \mathcal{L}[N]$ with $\eta=\begin{pmatrix} 1&0\\0&-1\end{pmatrix}$, using a block notation and
the hermitian scalar product $\langle , \rangle$ generating the norm $\Vert\ \Vert$:
\begin{equation}
\check{\mathcal{T}}_2[N](\Lambda,\Lambda^*)=\frac{1}{2} \meanvlr{ \begin{pmatrix} \Lambda,\Lambda^* \end{pmatrix},\eta \mathcal{L}[N] \begin{pmatrix} \Lambda \\ \Lambda^*\end{pmatrix} }
\label{eq:rond}
\end{equation}
Then one inserts the modal decomposition (\ref{eq:doubleetoile}) in the derivative with respect to $N$ of the equation (\ref{eq:rond});
we recall that $\epsilon_\alpha$ and $(u_\alpha,v_\alpha)$, $-\epsilon_\alpha$ and $(v_\alpha^*,u_\alpha^*)$ are the eigenvalues
and the eigenvectors of $\mathcal{L}$.  The coarse grained temporal average $\overline{\phantom{llll}}^t$ {removes} the crossed
terms, and the Hellmann-Feynman theorem ensures that\footnote{{\gilles $(u_\alpha,-v_\alpha)$ is the dual vector of $(u_\alpha,v_\alpha)$ 
because $(\eta \mathcal{L})^\dagger=\eta \mathcal{L}$.}} $\meanv{(u_\alpha,-v_\alpha) , \frac{\dd \mathcal{L}[N]}{\dd N} 
\begin{pmatrix} u_\alpha \\ v_\alpha \end{pmatrix}}=\dd \epsilon_\alpha / \dd N$, and finally leads to equation (\ref{eq:bla}).

\section*{Appendix B.\ \ Irrotational quantum hydrodynamics}


{In this supplementary section we give an additional derivation of equation (\ref{eq:herisson}) based on quantum hydrodynamics. Although this derivation is not microscopic contrarily to the one presented in the {\yvan main text of the paper and in appendix A}, and although it neglects the internal fermionic degrees of freedom treating the pairs of fermions at large spatial scales as a bosonic field, it has the advantage of relying only on the equation of state and thus of being valid in all interaction regimes.}

To calculate the viscosity of superfluid helium at low temperature, Landau and Khalatnikov have developed
in 1949 the theory of quantum hydrodynamics  \cite{Khalatnikov1949}. It allows one
to determine, to leading order in $T$, the effect of a non-zero temperature on the quantum fluid,
at least on the observables that only involve low energy scales and large length scales.
Remarkably, the only specific ingredient is the zero-temperature equation of state of the fluid, which is here
the energy per unit volume $e_0(\rho)$ of the ground state of the spatially homogeneous system of density $\rho$.

To obtain the time derivative of the phase operator of the condensate of fermion pairs, 
we refine the theory in two ways: by regularizing ultraviolet divergences and by specializing to the 
irrotational case.

First, we solve the issue of the Landau-Khalatnikov Hamiltonian ground state energy, that diverges due to
the zero-point motion of the system eigenmodes. We discretize the space in a cubic lattice of spacing ${l}$,
a sub-multiple of the size $L$ of the quantization volume, which is much smaller than the typical wavelength
$2\pi/q_{\rm th}$ of the thermal excitations of the fluid but much larger than the mean interparticle
distance $\rho^{-1/3}$,
\begin{equation}
\rho^{-1/3} \ll {l}\ll q_{\rm th}^{-1}
\label{eq:choixb}
\end{equation}
both conditions being compatible at sufficiently low temperature. This is in the spirit of the validity
range of hydrodynamics, which relies on a spatial coarse graining, and it provides a natural cut-off
for the wave vectors $\qq$ by restricting them to the first Brillouin zone\footnote{{\gilles We also require that
the elementary excitations of the system remain phononic over the whole domain $\mathcal{D}$.
This imposes $\hbar c/l<\Delta$ so $mcl/\hbar>mc^2/\Delta \gg 1$ in the BCS limit, and only $mcl/\hbar>1$ elsewhere, where $c$ 
is the sound velocity and $\Delta$ the gap at $T=0$. The necessary condition
$mc{l}/\hbar>1$ is weaker than the already 
assumed one, $\rho^{1/3}l\gg 1$, in the strongly interacting regime or in the BCS limit,
since $c$ is then of the order of the Fermi velocity. It is more stringent and must be added
explicitly in the so-called BEC limit, where the fermion pairs can be considered as bosons and form a weakly interacting condensate.}}
$\mathcal{D}=[-\pi/{l},\pi/{l}[^3$.   \setcounter{notecond}{\thefootnote}
In the Hamiltonian one must then replace the differential operators such as the gradient, the divergence
and the Laplacian, by their discrete versions, as we shall implicitly do below, and
introduce the {\sl bare} energy density $e_{0,0}(\rho)$, which depends on the lattice spacing ${l}$.
Following the ideas of renormalization, the zero-point energy of the modes, that formally diverges
when ${l}\to 0$, adds up to $e_{0,0}(\rho)$ to exactly reconstruct the {\sl effective } or {\sl true} 
energy density $e_0(\rho)$, that does not depend on ${l}$ and is what is measured experimentally.

Second, we specialize the theory to the case of an irrotational velocity field operator $\hat{\vv}(\rr,t)$ that
can then be written as the gradient of the phase field operator $\hat{\phi}(\rr,t)$, itself 
canonically conjugate to the density field operator $\hat{\rho}(\rr,t)$:
\begin{equation}
\hat{\vv}(\rr,t)=\frac{\hbar}{m}\mathrm{grad}\, \hat{\phi}(\rr,t)
\ \ \mbox{with}\ \ [\hat{\rho}(\rr,t),\hat{\phi}(\rr',t)]=\ii\frac{\delta_{\rr,\rr'}}{{l}^3}
\end{equation}
This amounts to neglecting the transverse component of the field $\hat{\vv}(\rr,t)$,
as done in reference  \cite{Khalatnikov1949} to determine the phonon-phonon interaction and go forward in the computation of viscosity.
In the particular scale invariant case of the unitary Fermi gas, this was justified within the effective field theory 
in reference  \cite{SonWingate2006}.
We note {\sl en passant} that the density of fermionic quasiparticles is exponentially small in $1/T$
and is directly omitted by hydrodynamic theory.

The steps that follow are rather usual. One starts form the equations of motion of the fields
in Heisenberg picture, that is the quantum continuity equation and the quantum Euler equation for the potential
(whose gradient gives the quantum Euler equation for the velocity):
\begin{eqnarray}
\label{eq:eq1}
\partial_t \hat{\rho} +\mathrm{div}\, \left[\frac{1}{2} \{\hat{\rho},\hat{\vv}\} \right] =0 \\
\hbar\partial_t \hat{\phi} = -\frac{1}{2} m \hat{\vv}^2 - \mu_{0,0}(\hat{\rho})
\label{eq:eq2}
\end{eqnarray}
where $\{\hat{A},\hat{B}\}=\hat{A} \hat{B} + \hat{B} \hat{A}$ is the anticommutator of two operators and 
\begin{equation}
\mu_{0,0}(\rho) = \frac{\mathrm{d}}{\mathrm{d}\rho} e_{0,0}(\rho) = e_{0,0}'(\rho)
\end{equation}
is the bare ground state chemical potential at density $\rho$.
{These equations originate from the Hamiltonian
\begin{equation}
\hat{H}_{\rm hydro} = l^3\sum_{\rr} \left[\frac{1}{2} m \hat{\vv}\cdot \hat{\rho} \hat{\vv}
+ e_{0,0}(\hat{\rho})\right]
\label{eq:Hamil_hydro}
\end{equation}
}
The quantum spatial density and phase fluctuations are weak provided that $\rho^{1/3} {l}$ is large enough;
the thermal ones are weak if in addition $q_{\rm th} {l}$ is small enough.\footnote{One first checks that the density fluctuations on a given lattice site are small in relative value, using
(\ref{eq:devmod1}). At $T=0$, $\langle \delta\hat{\rho}^2\rangle/\rho^2\approx \frac{\hbar}{mc{l}} \frac{1}{\rho {l}^3}\ll 1$, 
using (\ref{eq:choixb}) and the {\gilles necessary condition $mcl/\hbar >1$ established in} note \thenotecond. At $T>0$, there is the additional thermal contribution
$\approx\frac{\hbar}{mc{l}} \frac{(q_{\rm th} {l})^4}{\rho {l}^3}$
which is $\ll 1$ for the same reasons. Second, one checks that the phase fluctuations between two neighboring
lattice sites are small in absolute value. To this end, one notes from (\ref{eq:devmod2}) that 
${l}^2 \langle (\mathbf{grad}\, \delta{\hat{\phi}})^2\rangle =(mc{l}/\hbar)^2
\langle \delta\hat{\rho}^2\rangle/\rho^2$. To conclude, it remains to use $mc/(\hbar \rho^{1/3})=O(1)$, 
a property that holds in the whole BEC-BCS crossover, as well as the previous estimates 
of $\langle \delta\hat{\rho}^2\rangle/\rho^2$.}
Under these conditions one can linearize as in  \cite{Khalatnikov1949} the equations of motion around the spatially uniform solution:
\begin{eqnarray}
\label{eq:fluc1}
\hat{\rho}(\rr,t) & =&  \hat{\rho}_0 +\delta{\hat{\rho}}(\rr,t)  \\
\label{eq:fluc2}
\hat{\phi}(\rr,t) & =&  \hat{\phi}_0(t) +\delta{\hat{\phi}}(\rr,t)
\end{eqnarray}
The operator $\hat{\rho}_0$ reduces to $\hat{N}/L^3$, where $\hat{N}$ is the operator giving
the total number of particles, and is a constant of motion. The operator $\hat{\phi}_0$ is the phase 
operator of the condensate; one has here
\begin{equation}
\hat{\phi}_0=\hat{\theta}_0/2
\end{equation}
since the phase operator $\hat{\theta}_0$ in equation (\ref{eq:deftheta0}) takes the pairs as the building block,
whereas equations (\ref{eq:eq1},\ref{eq:eq2}) are build on the fermionic particles.
The spatial fluctuations $\delta{\hat{\rho}}$ and $\delta{\hat{\phi}}$, of vanishing (discrete) integral
over the whole space, can be expanded on the plane waves of non-zero wave vector $\qq$, and
commute with $\hat{\rho}_0$. One solves the linearized equations for $\delta\hat{\rho}$ and
$\delta{\hat{\phi}}$ and one can use the usual expansion on eigenmodes:
\begin{eqnarray}
\label{eq:devmod1}
\delta{\hat{\rho}}(\rr,t)=\frac{\hat{\rho}_0^{1/2}}{L^{3/2}}  \!\!
\sum_{\qq \in \frac{2\pi}{L}\mathbb{Z}^{3*}\cap \mathcal{D}} \!\!
\left(\frac{\hbar q}{2 m \hat{c}_{0,0}}\right)^{1/2} \!\! (\hat{B}_\qq+\hat{B}_{-\qq}^\dagger) \,\mathrm{e}^{\ii \qq\cdot \rr} \\
\label{eq:devmod2}
\delta{\hat{\phi}}(\rr,t)=\frac{-\ii}{\hat{\rho}_0^{1/2} L^{3/2}} \!\!
\sum_{\qq \in \frac{2\pi}{L}\mathbb{Z}^{3*}\cap \mathcal{D}} \!\!
\left(\frac{m \hat{c}_{0,0}}{2\hbar q}\right)^{1/2} \!\! (\hat{B}_\qq-\hat{B}_{-\qq}^\dagger) \,\mathrm{e}^{\ii \qq\cdot \rr} 
\end{eqnarray}
where the creation operator $\hat{B}^\dagger_\qq$ and the annihilation operator $\hat{B}_\qq$ of a phonon 
with wave vector $\qq$ and energy $\hbar q \hat{c}_{0,0}$ obey bosonic commutation relations
$[\hat{B}_\qq,\hat{B}_{\qq'}^\dagger]=\delta_{\qq,\qq'}$ and where we introduced the zero-temperature
bare sound velocity operator 
\begin{equation}
\hat{c}_{0,0} \equiv \left(\frac{\hat{\rho}_0 \mu_{0,0}'(\hat{\rho}_0)}{m}\right)^{1/2}
\end{equation}
It remains to expand the right-hand side of (\ref{eq:eq2}) up to second order in $\delta{\hat{\rho}}$
and $\delta{\hat{\phi}}$, to extract the zero wave vector Fourier component, to perform a coarse grained
temporal average to get rid of the oscillating crossed terms $\hat{B}_\qq \hat{B}_{-\qq}$ and
$\hat{B}^\dagger_{-\qq} \hat{B}^\dagger_{\qq}$, and to use the identity
\begin{equation}
\frac{\mathrm{d}}{\mathrm{d}\rho} [\rho \mu_{0,0}'(\rho)]^{1/2} = 
\frac{\mu_{0,0}'(\rho) + \rho \mu_{0,0}''(\rho)}
{2[\rho \mu_{0,0}'(\rho)]^{1/2}}
\end{equation}
to obtain
\begin{equation}
\hbar \overline{\frac{\mathrm{d}}{\mathrm{d}t} \hat{\phi}_0}^{t} = -\mu_{0,0}(\hat{\rho}_0)
-\!\!\!\!\sum_{\qq \in \frac{2\pi}{L}\mathbb{Z}^{3*}\cap \mathcal{D}}\!\!
\left(\hbar q \frac{\mathrm{d}}{\mathrm{d}\hat{N}} \hat{c}_{0,0}\right) 
\left(\hat{B}_\qq^\dagger \hat{B}_\qq +\frac{1}{2}\right)
\label{eq:but}
\end{equation}
At this order of the expansion, one can collect in (\ref{eq:but}) the zero-point contribution
of the modes (the term $1/2$ in between parentheses) and the bare chemical potential $\mu_{0,0}(\hat{\rho}_0)$
to form the true chemical potential $\mu_0(\hat{\rho}_0)$ of the fluid at zero temperature, and
one can identify $\hat{c}_{0,0}$ in the prefactor of the phonon number operator $\hat{B}_\qq^\dagger \hat{B}_\qq$
with the true sound velocity  at zero temperature, $\hat{c}_0\equiv[\hat{\rho}_0 \mu_0'(\hat{\rho}_0)/m]^{1/2}$.
One then obtains  the (low temperature) phononic limit of relation (\ref{eq:herisson}),
without any constraint on the interaction strength.

\providecommand*\hyphen{-}


\begin{thebibliography}{99}

\bibitem{Bloch2002_Revival}
M.~Greiner, O.~Mandel, T.~W. H\"ansch, I.~Bloch, 
Nature {419} 
(2002) 51.

\bibitem{Oberthaler2008}
J.~Est\`eve, C.~Gross, A.~Weller, S.~Giovanazzi, M.~K. Oberthaler, 
Nature {455} 
(2008) 1216.

\bibitem{Treutlein2010}
M.~F. Riedel, P.~B{\"o}hi, Yun Li, T.~W. H{\"a}nsch, A.~Sinatra, P. Treutlein,
Nature {464} 
(2010) 1170.

\bibitem{Ketterle2007}
G.-B. Jo, Y.~Shin, S.~Will, T.~A. Pasquini, M.~Saba, W. Ketterle, D.~E. Pritchard, M. Vengalattore, M. Prentiss,
Phys. Rev. Lett. {98} (2007) 030407.

\bibitem{Sidorov2011}
M.~Egorov, R.~P. Anderson, V.~Ivannikov, B.~Opanchuk, P.~Drummond, B.~V. Hall, A.~I. Sidorov,
Phys. Rev. A {84} (2011) 021605.

\bibitem{Berrada2013}
T.~Berrada, S.~van Frank, R.~B\"{u}cker, T.~Schumm, J.-F. Schaff, J. Schmiedmayer,
Nature Communications {4} (2013) 2077.

\bibitem{Salomon2010}
S.~Nascimb{\`e}ne, N.~Navon, K.~J. Jiang, F.~Chevy, C.~Salomon, 
Nature {463} 
(2010) 1057.

\bibitem{Zwierlein2012}
M.~J.~H. Ku, A.~T. Sommer, L.~W. Cheuk, M.~W. Zwierlein, 
Science {335} 
(2012) 563.

\bibitem{Kohstall2011}
C.~Kohstall, S.~Riedl, E.~R. {S\'{a}nchez Guajardo}, L.~A. Sidorenkov,  J.~{Hecker Denschlag}, R. Grimm, 
New Journal of Physics {13} 
(2011) 065027.

\bibitem{CarusottoCastin2005}
I.~Carusotto, Y.~Castin, 
Phys. Rev. Lett. {94} (2005) 223202.

\bibitem{PRA2013}
H.~Kurkjian, Y.~Castin, A.~Sinatra, 
Phys. Rev. A {88} (2013) 063623.

\bibitem{Sinatra2007}
A.~Sinatra, Y.~Castin, E.~Witkowska, 
Phys. Rev. A {75} (2007) 033616.

\bibitem{Kuklov2000}
A.~B. Kuklov, J.~L. Birman, 
Phys. Rev. A {63} (2000) 013609.

\bibitem{Anderson1958}
P.~Anderson, 
Phys. Rev. {112} (1958) 1900.

\bibitem{Cohen}
C.~Cohen-Tannoudji, J.~Dupont-Roc, G.~Grynberg, Processus d'interaction entre
  photons et atomes, InterEditions et \'Editions du {CNRS}, Paris (1988).

\bibitem{Leggett2006}
A.~J. Leggett, {Quantum Liquids}, Oxford University Press, Oxford (2006).

\bibitem{Olshanii2008}
M.~Rigol, V.~Dunjko, M.~Olshanii, 
Nature {452} 
(2008) 854.

\bibitem{WitkowskaCastinSinatra2009}
A.~Sinatra, Y.~Castin, E.~Witkowska, 
Phys. Rev. A {80} (2009) 033614.

\bibitem{Ripka1985}
J.-P. Blaizot, G.~Ripka, Quantum {T}heory of {F}inite {S}ystems, MIT Press, Cambridge, Massachusetts (1985).

\bibitem{CastinDum1998}
Y.~Castin, R.~Dum, 
Phys. Rev. A {57} (1998) 3008.

\bibitem{CKS2006}
R.~Combescot, M.~Yu. Kagan, S.~Stringari, 
Phys. Rev. A {74} (2006) 042717.

\bibitem{Khalatnikov1949}
L.~Landau, I.~Khalatnikov, 
Zh. Eksp. Teor. Fiz.  {19} (1949) 637.

\bibitem{Grimm2004}
M.~Bartenstein, A.~Altmeyer, S.~Riedl, S.~Jochim, C.~Chin, J. Hecker Denschlag, R. Grimm, 
Phys. Rev. Lett. {92} (2004) 120401.

\bibitem{Bloch2002}
M.~Greiner, O.~Mandel, T.~Esslinger, T.~W. H\"ansch, I.~Bloch, 
Nature {415} 
(2002) 39.

\bibitem{Bakr2009}
W.~S. Bakr, J.~I. Gillen, A.~Peng, S.~F\"olling, M.~Greiner, 
Nature {462} 
(2009) 74.

\bibitem{CastinSinatra2012}
Y.~Castin, A.~Sinatra, 
chapter \g{Spatial and Temporal Coherence of a Bose-condensed Gas}
dans Physics of Quantum fluids: New Trends and Hot Topics in Atomic and Polariton Condensates, 
A.~Bramati, M.~Modugno (r\'edacteurs), Springer, Berlin (2013).

\bibitem{Hadzibabic2013}
A.~L. Gaunt, T.~F. Schmidutz, I.~Gotlibovych, R.~P. Smith, Z.~Hadzibabic,
Phys. Rev.  Lett. {110} (2013) 200406.

\bibitem{Bulgac2006}
A.~Bulgac, J.~E. Drut, P.~Magierski, 
Phys. Rev. Lett. {96} (2006) 090404.

\bibitem{Son2006}
Y.~Nishida, D.~T. Son, 
Phys. Rev. Lett. {97} (2006) 050403.

\bibitem{Ketterle2008}
A.~Schirotzek, Yong-il Shin, C.~H. Schunck, W.~Ketterle, 
Phys.  Rev. Lett. {101} (2008) 140403.

\bibitem{Castin2015}
Y.~Castin, I.~Ferrier-Barbut, C.~Salomon, 
C. R. Physique {16} (2015) 241.

\bibitem{Salasnich2015}
G.~Bighin, L.~Salasnich, P.~A. Marchetti, F.~Toigo, 
Phys. Rev. A {92} (2015) 023638.

\bibitem{Tempere2011}
S.~N. Klimin, J. Tempere, Jeroen P.~A. Devreese, J. Low. Temp. Phys. 165 (2011) 261.

\bibitem{KCS2016}
H. Kurkjian, Y. Castin, A. Sinatra, Phys. Rev. A 93 (2016) 013623.

\bibitem{SonWingate2006}
D.~T.~ Son, M.~Wingate, 
Ann. Physics {321} (2006) 197.

\end{thebibliography}
\end{document}